\documentstyle[12pt,twoside]{article}
\begin{document}
\begin{center}
{\Large\bf QUANTUM COSMOLOGICAL PERFECT\\[5pt] FLUID MODEL\\[5PT]
AND ITS CLASSICAL ANALOGUE\\[5PT]}\medskip 
{\bf A.B. Batista$^\dag$\,\footnote{e-mail: brasil@cce.ufes.br},
J.C. Fabris$^\dag$\,\footnote{e-mail: fabris@cce.ufes.br}, 
S.V.B. Gon\c{c}alves$^\dag$\,\footnote{e-mail: sergio@cce.ufes.br}
and J. Tossa$^{\dag\ddag}$\,\footnote{e-mail:jtoss@syfed.bj.refer.org}} 
\medskip
\\
\dag Departamento de F\'{\i}sica, 
Universidade Federal do Esp\'{\i}rito Santo, 29060-900, Vit\'oria, 
Esp\'{\i}rito Santo, Brazil \medskip\\
\ddag IMSP - Universit\'e Nationaledu B\'enin, Porto Novo, B\'enin\medskip\\
\medskip\end{center} 
\begin{abstract}

The quantization of gravity coupled to a perfect fluid model leads to a
Schr\"odinger-like equation, where the matter variable plays the role of
time. The wave function can be determined, in the flat case, for an arbitrary barotropic
equation of state $p = \alpha\rho$; solutions can also be found for the radiative non-flat case. The wave packets are constructed, from
which the expectation
value for the scale factor is determined. The quantum scenarios reveal
a bouncing Universe, free from singularity. We show that such quantum
cosmological perfect fluid models admit a universal classical analogue,
represented by the addition, to the ordinary classical model, of a repulsive
stiff matter fluid. The meaning of the existence of this universal
classical analogue is discussed. The
quantum cosmological perfect fluid model is, for a flat spatial section,
formally equivalent to a free particle in ordinary quantum mechanics,
for any value of $\alpha$, while the radiative non-flat case is equivalent
to the harmonic oscillator. The repulsive fluid needed to
reproduce the quantum results is the same in both cases.\vspace{0.7cm}PACS number(s): 04.20.Cv., 04.20.Me\end{abstract}
\section{Introduction}

The standard cosmological model predicts that the Universe had an initial
singular state, from which expansion followed. In spite of many success of
this scenario, it is hard to believe that such initial state may have existed,
since it is not possible to construct a physical scenario for a singular
state.
It is generally argued that quantum effects appear as the Universe approaches
the initial singularity, the temperature mounting to levels comparable to
the Planck temperature, leading to the avoidance of that singular initial state.
The great problem with this mechanism is the absence, until today, of a consistent
quantum theory of gravity. However, quantum cosmology permits, in principle, to
circumvent this difficulty, since it signs with the possibility that the entire Universe
may admit a quantization procedure, from which a quantum scenario can be built up.
\par
However, quantum cosmology faces many conceptual and technical problems \cite{halliwell,nelson,isham1,isham2}. First,
since it is based on the ADM decomposition, leading to the hamiltonian formulation
of General Relativity, it can be applied only to space-times admiting foliation.
From the cosmological point of view, this does not seem to be a serious
restriction. However, the equation for the so-called wave function of the Universe,
the Wheeler-DeWitt equation, is a functional equation defined in the superspace,
the space of all possible spatial geometries, and no solution for it is known until
now, unless an infinite number of degrees of freedom is frozen. Moreover, the gravity theory reveals to be a constrained system, and the
time reparametrisation freedom of General Relativity implies that the
superhamiltonian of gravity is zero. The consequence is that, in the process of
quantization, time disappears.
\par
In some situations, the notion of time can be recovered in quantum cosmology.
One example is the case where gravity is coupled to a perfect fluid.
Employing the Schutz's formalism for the description of the perfect fluid \cite{schutz1,schutz2},
based on some auxiliary potentials which represent the dynamical degrees of
freedom of the fluid, we can obtain a hamiltonian in the minisuperspace,
where the scale factor and the fluid variable are the only dynamical degrees of freedom. 
The conjugate momentum
associated to the variables of the fluid appears linearly. After quantization,
this implies a Schr\"odinger equation, where this conjugate momentum plays the role
of time. The solution of this Schr\"odinger equation leads to eigenfuctions which
are not square integrable. But finite wave functions can be constructed by superposing
these eigenfunctions conveniently. The expectation value for the dynamical variable
(in this case, the scale factor of the Universe) can be obtained, revealing a
singularity-free Universe which exhibit a bounce and approaches asymptoticaly
the classical solution.
\par
The purpose of the present paper is to show that such quantum perfect fluid model
can be mimitized, from the point of view of the behaviour of the
scale factor, by a classical model where, besides the ordinary fluid characterized by
the equation of state $p = \alpha\rho$,
a repulsive fluid with an equation of state $p = \rho$ is introduced, which does not depend on
the value of $\alpha$ and, perhaps, on the spatial curvature $k$. This generalizes the results obtained in
\cite{brasil}. In principle, a suitable classical model can reproduce certain
aspects of a given quantum model. What is surprising here is that the classical
model which reproduces the quantum cosmological perfect fluid model is
obtained by introducing a universal repulsive fluid: the analogous classical model is valid
for the flat case, which is formally equivalent to the free particle problem in
ordinary quantum mechanics, and for the radiative curved case, which takes the
form of a harmonic oscillator problem. We guess that other quantum cases, different from
those quoted before, may also be reproduced by the analogous classical model.
\par
The existence of such analogous classical model to the quantum cosmological perfect
fluid model in so many different situations may rise the question if the quantization of a perfect fluid, at least
in the mini-superspace, is a real quantization. We do not intend to answer this question
here. But we notice that the important difference between the cases treated here and the
equivalent cases in ordinary quantum mechanics (free particle, harmonic oscillator) is
connected to the fact that the dynamical variable in the present problem (the scale factor of
the Universe) is positive definite, while in the quantum ordinary case the position variable
is definite in all real axis. This restriction to half of the real axis is imposed by
the fact that the system is coupled to gravity. So the classical analogue exhibited here
does not occur in ordinary quantum mechanics, being specific of the quantum cosmological
perfect fluid model.
\par
This paper is organized as follows. In next section, we describe the quantum model,
with the solutions for the flat case, in the minisuperspace. Wave packets are constructed, and
the expectation value of the scale factor is computed. In section 3, the spreading
of the wave packet is analyzed, revealing a behaviour very close to the free particle
problem in ordinary quantum mechanics. The similarity of the flat quantum perfect fluid model with the
free particle problem in ordinary quantum mechanics is discussed. In section 4, the classical analogue is presented.
The problem with curvature in the spatial section is discussed in section 5, and
it is shown the classical analogue can still be applied, at least for a radiative fluid.
Section 6 contains our conclusions.

\section{The quantum model}

The action for a perfect fluid coupled to gravity in the Schutz's formalism may be written
as
\begin{equation}
\label{acao}
S = \int_M d^4x\sqrt{-g}p - \int_Md^4x\sqrt{-g}R - \int_{\partial M}d^3x\sqrt{h}K \quad ,
\end{equation}
where $p$ is the pressure, $K$ is the trace of the extrinsic curvature and $h$ is the
induced metric in the boundary of the manifold $M$.
The four velocity of the fluid is written with the aid of five potentials $\phi$, $\epsilon$,
$\beta$, $\theta$ and $S$:
\begin{equation}
u_\nu = \frac{1}{\mu}(\epsilon_{,\nu} + \phi\beta_{,\nu} + \theta S_{,\nu})
\end{equation}
where $\mu$ is the specific enthalpy. The four velocity is subjected to the
condition $u^\nu u_\nu = 1$.
Introducing the Friedmann-Lema\^{\i}tre-Robertson-Walker metric in the
action (\ref{acao}), it is possible to construct the super-hamiltonian \cite{nivaldo}
\begin{equation}
H = - \frac{\Pi_a^2}{24a} - 6ka + \frac{p_\epsilon^{1+\alpha}e^S}{a^{3\alpha}} \quad ,
\end{equation}
where $\alpha$ defines de equation of state of the fluid ($p = \alpha\rho$).
Performing the canonical transformations
\begin{equation}
T = p_Se^{-S}p_\epsilon^{-(1+\alpha)} \quad , \quad \Pi_T = e^Sp_\epsilon^{1+\alpha} \quad .
\end{equation}
we finally obtain the following expression for the super-hamiltonian:
\begin{equation}
\label{hpf}
H = - \frac{\Pi_a^2}{24a} - 6ka + \frac{\Pi_T}{a^{3\alpha}} 
\end{equation}
where
$\Pi_T$ is the momentum associated with the matter variable. It appears linearly
in the hamiltonian. The parameter $k$ defines the curvature of the
spatial section, taking the values 0, 1, - 1 for a flat, closed and open
Universe, as usual. The Schutz's formalism for the description of perfect fluids
gives dynamical degrees of freedom to them. Hence, the hamiltonian constraint
can not be used to eliminate all the degrees of freedom of the problem.
\par
Imposing the quantization conditions and aplying this hamiltonian to
the wave function, we obtain the Wheeler-DeWitt equation in the
minisuperspace:
\begin{equation}
\label{se}
\frac{\partial^2\Psi}{\partial a^2} - 144ka^2 - i24a^{1 - 3\alpha}\frac{\partial\Psi}{\partial T} = 0 \quad .
\end{equation}
Due to the canonical transformations employed in the construction of the hamiltonian,
the time coordinate $T$ is connected with the cosmic time $t$ by the relation
$dt = a^{3\alpha}dT$.
Notice that the equation (\ref{se}) is equivalent to the Schr\"odinger equation of ordinary quantum
mechanics. Hence, all the formalism of quantum mechanics, like Hilbert space, observables
represented by self-adjoint operators, can be applied to this problem.
In order the hamiltonian operator to be self-adjoint, the scalar product between
the wave functions $\Phi$ and $\Psi$ must take the form
\begin{equation}
\label{scalar}
(\Phi,\Psi) = \int_0^\infty a^{1 - 3\alpha}\Phi^*\Psi da \quad .
\end{equation}
The non-conventional measure in the scalar product (\ref{scalar}) implies that $a$ must
be positive definite in order to have a positive norm for the wave function, except for
some specific equation of state, e.g., those for which $\alpha = 1/3, -1/3, -1$.
However, the physical requirement that $a$ must specify the scale, force us to restrict
it to positive values, even for those particular cases.
\par
The Wheeler-DeWitt equation written above may be solved through the
separation of variables method. Indeed, writing
\begin{equation}
\Psi(a,T) = e^{iET}\xi(a)
\end{equation}
we obtain
\begin{equation}
\label{de1}
\xi'' + \biggr(-144ka^2 + 24Ea^{1 - 3\alpha}\biggl)\xi = 0 \quad ,
\end{equation}
where the primes mean derivative with respect to $a$. It is possible
to show that the parameter $E$ is positive, except when $\alpha = 1$,
a special case which will be discussed separately later.
Notice that in principle the order ambiguity in (\ref{se}) should be taken into account.
But, it is possible to show that the final results do not depend on it.
\par
When $k = 0$, the equation (\ref{de1}) admits a solution under the form of
Bessel functions, leading to the following final expression for
the wave function:
\begin{equation}
\Psi = e^{iET}\sqrt{a}\biggr[c_1J_{\frac{1}{3(1 - \alpha)}}\biggr(\frac{\sqrt{96E}}{3(1 - \alpha)}a^{\frac{3(1 - \alpha)}{2}}\biggl) + c_2J_{-\frac{1}{3(1 - \alpha)}}\biggr(\frac{\sqrt{96E}}{3(1 - \alpha)}a^{\frac{3(1 - \alpha)}{2}}\biggl)\biggl] \quad .
\end{equation}
These solutions are not valid for $\alpha = 1$. In this particular case,
equation (\ref{de1}) becomes an Euler's type equation, and the
final solution takes the form:
\begin{equation}
\Psi = e^{iET}\sqrt{a}\biggr[c_1a^{\sqrt{1 - \frac{96E}{2}}} +
c_2a^{-\sqrt{1 - \frac{96E}{2}}}\biggl] \quad .
\end{equation}
\par
The above solutions must obey convenient boundary conditions in order the
original hamiltonian operator to be self-adjoint. These boundary conditions
are
\begin{equation}
\Psi(0,T) = 0 \quad \mbox{or} \quad \frac{\partial\Psi}{\partial a}(a,T)\vert_{a = 0} = 0 \quad .
\end{equation}
The first one amounts to impose $c_2 = 0$, while the second one
implies $c_1 = 0$. For $\alpha = 1$ it is possible to satisfy the boundary conditions
but the value of $E$ must be bounded from above. This compromises the possibility to
construct well-behaved wave packets. Hence this case seems to be pathological.
\par
None of the above solutions is square integrable. Hence, wave packets
must be constructed, by superposing those solutions, in order to obtain
expressions with physical meaning.
The general structure of these superpositions is
\begin{equation}
\Psi(a,T) = \int_0^\infty A(E)\Psi_E(a,T)dE \quad .
\end{equation}
We specialize the discussion from now on to the case $\alpha < 1$ and
$c_2 = 0$. Nothing would have changed if we have choosed $c_1 = 0$ instead.
Defining $r = \frac{\sqrt{96E}}{3(1 - \alpha)}$, we can obtain final
closed expressions for the wave packet if we choose the function
$A(E)$ to be a quasi-gaussian superposition factor:
\begin{equation}
\Psi(a,T) = \sqrt{a}\int_0^\infty r^{\nu + 1}e^{-\gamma r^2 + i\frac{3}{32}
(1 - \alpha)^2r^2T}J_\nu(ra^\frac{3(1 - \alpha)}{2})dr \quad ,
\end{equation}
where $\nu = \frac{1}{3(1 - \alpha)}$ and $\gamma$ is a real parameter.
The wave packet takes then the form
\begin{equation}
\label{packet}
\Psi(a,T) = a\frac{e^{-\frac{a^{3(1 - \alpha)}}{4B}}}{(-2B)^{\frac{4 - 3\alpha}{3(1 - \alpha)}}} \quad ,
\end{equation}
where $B = \gamma - i\frac{3}{32}(1 - \alpha)^2T$.
\par
Now, we are interested in verifying which are the previsions of those
quantum models for the behaviour of the scale factor of the Universe.
In order to do this, we adopt the many worlds interpretation \cite{tipler} and
calculate the expectation value of the scale factor:
\begin{equation}
<a>_T = \frac{\int_0^\infty a^{1 - 3\alpha}\Psi(a,T)^*a\Psi(a,T)da}
{\int_0^\infty a^{1 - 3\alpha}\Psi(a,T)^*\Psi(a,T)da} \quad .
\end{equation}
The integrals above are easily solved, leading to
\begin{equation}
\label{wp1}
<a>_T = \biggr(\frac{\Gamma[\frac{5 - 3\alpha}{3(1 - \alpha)}]}
{\Gamma[\frac{4 - 3\alpha}{3(1 - \alpha)}]}\biggl)
(2\gamma)^\frac{1}{3(1 - \alpha)}\biggr[\frac{9(1 - \alpha)^4}{(32)^2\gamma^2}T^2 + 1\biggl]^\frac{1}{3(1 - \alpha)} \quad .
\end{equation}
These solutions represent a bouncing Universe, with no singularity,
which goes asymptotically to the corresponding classical model
when $T \rightarrow \infty$. They were first written down in \cite{nivaldo}.

\section{Spreading of the wave packet}

In ordinary quantum mechanics, the width of the wave packet permits
to have important informations about the classical limit of the
quantum model. This width is defined as
\begin{equation}
\Delta^2 = <a^2> - <a>^2 \quad .
\end{equation}
In the case of a free particle, for example, the distribution of the
probability of finding the particle coincides asymptotically for
large values of time with the distribution of
trajectories of classical particles which had initially some spread in
their initial conditions in the origin.
Of course, this does not mean that the notion
of classical trajectory is recovered asymptotically.
As the gaussian factor goes to zero, the particle is completly localized
and the notion of classical trajectory is recovered.
\par
In quantum cosmology the situation is more subtle. At the present stage, it seems that there exist
two main interpretation schemes that are specially usefull in quantum
cosmology: the many worlds interpretation and the Bohm-de Broglie interpretation.
In the last one, the notion of trajectory is essential. In the former one,
this question is less clear but it can be considered that it predicts
also trajectories that bifurcate following the possible eigenvalues, leading
to what is called consistent histories. Hence, it seems that it is very
difficult to avoid the notion of trajectories in quantum cosmologies,
due to the unicity of the Universe, even if in the many worlds interpretation
we have many possible Universes that do not communicate among themselves.
\par
Let us evaluate the spread of the wave packet determined before.
Using the expression (\ref{packet}), we find
\begin{eqnarray}
<a^2> &=& \frac{\int_0^\infty a^{1 - 3\alpha}\Psi^*(a,T)a^2\Psi(a,T)da}{\int_0^\infty a^{1 - 3\alpha}\Psi^*(a,T)\Psi(a,T)da}\\
&=& \frac{\Gamma[\frac{2 - \alpha}{1 - \alpha}]}{\Gamma[\frac{4 - 3\alpha}{3(1 - \alpha)}]}
\biggr\{\frac{2\gamma^2 + \frac{18}{(32)^2}(1 - \alpha)^4T^2}{\gamma}\biggl\}^\frac{2}{3(1 - \alpha)} \quad .
\end{eqnarray}
The expression for the width of the wave packet takes then the
form
\begin{equation}
\label{swp1}
\Delta^2 = \frac{\Gamma[\frac{2 - \alpha}{1 - \alpha}] 
- \Gamma^2[\frac{5 - 3\alpha}{3(1 - \alpha)}]}{\Gamma^2[\frac{4 - 3\alpha}{3(1 - \alpha)}]}\biggr\{\frac{2\gamma^2 + \frac{18}{(32)^2}(1 - \alpha)^4T^2}{\gamma}\biggl\}^\frac{2}{3(1 - \alpha)} \quad .
\end{equation}
\par
There is a striking similarity between the result for the spreading of
the wave packet found here and the case of the free particle in ordinary
quantum mechanics \cite{cohen}. This similarity will be discussed in more details
later in this section.
Considering an ensemble of Universe with some
spreading in their initial conditions, then the classical trajectories
coincide with the quantum "trajectories" asymptotically for large values
of time. The classical solution is recovered from the
expectation value of the quantum solutions, for any value of time,
when the gaussian parameter $\gamma$ goes to zero. Moreover, the
wave packet has an initial small width that spreads as time evolves.
\par
Even if technically the results of the quantum model for the Universe
with a perfect fluid is similar to the result for a free particle,
the question of interpretation makes all the difference. In the case of a
free particle in ordinary quantum mechanics,
the notion of trajectory is not recovered asymptotically.
However, in quantum cosmology it seems unavoidable that the Universe
must follows a trajectory. Hence, if we
adopt this point of view, the results exhibited above tell us that
the ensemble of "quantum trajectories" initially differs from the
possible classical trajectories,
coinciding with them later. The initial discrepancy is due to the appearence of
repulsive quantum effects as the existence of a classical analogue will
reveal explicitly.
\par
As it has already been remarked, the behaviour for the quantum perfect fluid model presented before has a great
resemblance with what happens in the free particle problem in ordinary quantum
mechanics. In this case, the solution of the Schr\"odinger equation leads to
the wave function
\begin{equation}
\Psi(x,t) = A(k)\exp{i(kx - \omega t)} \quad ,
\end{equation}
with $\omega = \hbar k^2/2m$. This plane wave solution is not realistic, and the
wave function for a free particle is obtained by constructing the wave packet
\begin{equation}
\Psi(x,t) = \frac{1}{(2\pi)^\frac{3}{2}}\int A(k)\exp{i[kx - \omega(k)t]}dk \quad .
\end{equation}
Even if a gaussian superposition does not lead to an expression for the expectation value of the
position similar to (\ref{wp1}), the wave packet spreads as it evolves in a manner
similar to (\ref{swp1}).
This similarity in fact express a more deep connection between the quantum mechanical
free particle and the flat quantum cosmological perfect fluid model.
\par
In fact, let us consider again the super-hamiltonian (\ref{hpf}). From it, we can
define, for $k = 0$, the reduced hamiltonian
\begin{equation}
\label{rh}
H_r = \frac{1}{24}a^{3\alpha-1}p_a^2 \quad .
\end{equation}
It is this reduced hamiltonian that drives the evolution of the system, in the sense
of ordinary quantum mechanics. The Wheeler-DeWitt equation in the minisuperspace can
be written as a genuine Schr\"odinger equation,
\begin{equation}
H_r\Psi = i\frac{\partial\Psi}{\partial T} \quad .
\end{equation}
If a
canonical transformation such that
\begin{equation}
p_x = \frac{1}{\sqrt{12}}a^{(3\alpha-1)/2}p_a \quad , \quad x =  \frac{4}{\sqrt{3}}\frac{a^{(1 - 3\alpha)/2}}{1-\alpha} \quad ,
\end{equation}
is performed in (\ref{rh}), then the reduced hamiltonian takes the form
\begin{equation}
H_r = \frac{1}{2}p_x^2
\end{equation}
which is equivalent to the free particle problem in ordinary quantum mechanics.
In \cite{demaret} such identification was made for $\alpha = 0$.
Here, this reduction is valid for any value of $\alpha$.
To reduce the original problem to a free particle problem through a convenient canonical
transformation does not mean that their physical content is the same. For example, the harmonic
oscillator problem may also be expressed in terms of a free particle problem through
a canonical transformation \cite{goldstein}. But, their physical contents are very different. The physical interpretation must be made in the original
variables, which in the present case is the scale factor $a$. The
quantum perfect fluid model is equivalent to a free particle, strictly speaking,
only for $\alpha = 1/3$; however, for other values of $\alpha$ the free particle expression
may be obtained through simple redefinitions, even if original problem
is not exactly the free particle one. Another important difference is that, in
the quantum mechanics ordinary case $- \infty < x < \infty$, while in the quantum
cosmological case $0 \leq a < \infty$.

\section{The classical analogous model}

Now, a classical model which reproduces the results found before for the expectation
value for the scale factor is worked out. This classical analogue is obtained by
considering a Friedmann model with an ordinary perfect fluid model, with an
equation of state $p = \alpha\rho$  (the same employed in the quantization in
the previous sections), plus a repulsive fluid with an equation of state
$p_q = \rho_q$, where the subscript $q$ was chosen in order to remember that
we look for a term that may reproduce the quantum effects.
\par
The Einstein's field equations reduces then to
\begin{equation}
\label{ce}
\biggr(\frac{\dot a}{a}\biggl)^2 = \frac{C_1}{a^{3(1 + \alpha)}} -
\frac{C_2}{a^6} \quad .
\end{equation}
This equation may be solved by reparametrizing the time coordinate
as
\begin{equation}
dt = a^{3\alpha}dT \quad ,
\end{equation}
leading to the expression
\begin{equation}
\biggr(\frac{a'}{a}\biggl)^2 = C_1a^{-3(1 - \alpha)} - C_2a^{-6(1 - \alpha)} \quad .
\end{equation}
This equation can be easily solved, leading to the following expression
for the scale factor:
\begin{equation}
\label{cs}
a(T) = \biggr(\frac{C_1}{C_2}\biggl)^\frac{1}{3(1 - \alpha)}\biggr[\frac{{C_1}^2C_2}{36(1 - \alpha)^2}T^2 +
1\biggl]^\frac{1}{3(1 - \alpha)}
\quad .
\end{equation}
\par
The first thing to notice is that the time coordinate $T$ is the same obtained in the
quantum model: this is due to the choice of the canonical variable in the quantum
model.
Consequently, the solution (\ref{cs}) is essentially the same as that obtained for
the scale factor expectation value in the quantum model.
In fact, both solutions coincide quantitativelly if we fix the integration
constants as
\begin{equation}
C_1 = \biggr(\frac{\Gamma[\frac{5 - 3\alpha}{3(1 - \alpha)}]}
{\Gamma[\frac{4 - 3\alpha}{3(1 - \alpha)}]}\biggl)^{1 - \alpha}\frac{3}{8}
3^{1/3}\frac{(1 - \alpha)^2}{\gamma} \quad , \quad 
C_2 = \biggr(\frac{\Gamma[\frac{5 - 3\alpha}{3(1 - \alpha)}]}
{\Gamma[\frac{4 - 3\alpha}{3(1 - \alpha)}]}\biggl)^{-2(1 - \alpha)}\frac{3}{4}
3^{1/3}(1 - \alpha)^2 \quad .
\end{equation}
This remark is valid for any value of $\alpha \leq 1$, covering
all the "free" particle problem (we remember
that the free particle problem occurs, strictly speaking, only for $\alpha = 1/3$).
For $\alpha = 1$, the solution for (\ref{ce}) depends on the relative values of
$C_1$ and $C_2$: for $C_1 > C_2$, the traditional solution for a stiff matter
is obtained with $a \propto t^{1/3}$; for $C_1 = C_2$, the only solution
is the Minkowski space-time; for $C_1 < C_2$, there is no lorentzian solution.
This must be compared with the fact, stressed before, that there is no
consistent quantum
solution for this case. The introduction of a order factor in (\ref{se}) would just
change the argument of the gamma functions.
\par
The existence of a repulsive term implies that the energy conditions are
violated as the singularity is approached, leading to its avoidance.
In fact, let us consider the null energy condition
\begin{equation}
8\pi G(\rho + p) \geq 0 \quad .
\end{equation}
This condition establishes that a comoving observer measures a
positive energy density; if this condition is violated, the co-moving
observer will measure a negative energy density, and repulsive effects
take place.
Considering the $\rho_{eff}$ and $p_{eff}$ as the sum of the energy and pressure
for both the attractive and repulsive fluids,
we find
\begin{equation}
8\pi G(\rho_{eff} + p_{eff}) = \frac{2}{a^{6\alpha}}\biggr( - \frac{a''}{a} + 
(1 + 3\alpha)\frac{a'^2}{a^2}\biggl) \quad ,
\end{equation}
with primes meaning derivative with respect to $T$.
Inserting the solutions (\ref{cs}), with an unimportant absortion of
integration constant in the definition of the time coordinate,
we find
\begin{equation}
8\pi G(\rho_{eff} + p_{eff}) = \frac{1}{a^{6\alpha}}\frac{4}{3(1 - \alpha)^2}\biggr[\frac{(1 + \alpha)T^2
- (1 - \alpha)}{(T^2 + 1)^2}\biggl] \quad ,
\end{equation}
which becomes negative for $T < \sqrt{\frac{1 - \alpha}{1 + \alpha}}$.
Hence, for each value of $\alpha$ smaller than one, the dominant
energy condition is violated around the bounce. For negative values
of $\alpha$ the period of time during which the energy conditions are violated becomes
larger, and in particular for $\alpha = - 1$ the energy conditions
are violated for any value of time. 

\section{The radiative case with curvature}

If the curvature of the spatial section is taken into account,
the integration of the Wheeler-DeWitt equation in the minisuperspace
is not so easy, and perhaps there is no simple analytical solution
for any value of $\alpha$. However, for some values of $\alpha$ we can
integrate the equations, construct explicitly the wave packets and obtain
the expectation value for the scale factor as before.
\par
We will consider now the radiative case, for which such integration
is possible, being also a very important particular case. For the radiative case
the time coordinate $T$ becomes identical to the conformal time $\eta$.
The Wheeler-DeWitt equation in the minisuperspace may be written
as
\begin{equation}
\label{curv1}
\frac{\partial^2\Psi}{\partial a^2} - 144ka^2\Psi - i24\frac{\partial\Psi}{\partial\eta} = 0 \quad .
\end{equation}
Notice that the radiative case is equivalent to the quantum harmonic oscillator \cite{barrow}.
The analysis of this equation is more involved. In \cite{rubakov,nivaldo1,
nelson1} Green's function methods were employed, and gaussian superposition
were constructed.
\par
The equation (\ref{curv1}) can be solved by writing
\begin{equation}
\Psi(a,\eta) = \int_0^\infty G(a,a',\eta)\Psi_0(a')da' \quad .
\end{equation}
The function $\Psi_0(a)$ defines the initial configuration, which must
satisfy the boundary conditions specified before.
The propagator takes the form \cite{rubakov,nivaldo1}
\begin{equation}
G(a,a',\eta) = \sqrt{\frac{6\sqrt{k}}{i\pi\sin(\sqrt{k}\eta)}}
\exp\biggr\{\frac{6i\sqrt{k}}{\sin(\sqrt{k}\eta)}\biggr[(a^2 + a'^2)cos(\sqrt{k\eta}) - 2aa'\biggl]\biggl\} \quad .
\end{equation}
Choosing the initial configuration for the wave function as
\begin{equation}
\Psi_0 = \biggr(\frac{8\sigma}{\pi}\biggl)^{1/4}\exp(-\beta a^2) \quad , \quad \beta = \sigma + ip \quad ,
\end{equation}
$\sigma$ and $p$ being real parameters,
we obtain the following wave function for a curved radiative Universe:
\begin{eqnarray}
\Psi(a,\eta) = \biggr(\frac{8\sigma}{\pi}\biggl)^{1/4}
\biggr\{\frac{6\sqrt{k}}{\cos(\sqrt{k}\eta)[\beta\tan(\sqrt{k}\eta) -
6i\sqrt{k}]}\biggl\}^{1/2}\nonumber\\
\times\exp\biggr\{\frac{6i\sqrt{k}}{\tan(\sqrt{k}\eta)}\biggr(1 
+ \frac{6i\sqrt{k}}{\cos^2(\sqrt{k}\eta)[\beta\tan(\sqrt{k}\eta) -
6i\sqrt{k}]}\biggl)a^2\biggl\} \quad .
\end{eqnarray}
Calculating the expectation value for the scale factor as before,
we obtain
\begin{equation}
\label{rqs1}
<a>_\eta = \left\{\sqrt{\sigma^2\sin^2\eta + (6 - p\tan\eta)^2\cos^2\eta}
\quad \hfill\hbox{k = +1} \quad ,
\atop
\sqrt{\sigma^2\sinh^2\eta + (6 - p\tanh\eta)^2\cosh^2\eta} \quad \hbox{k = -1}\quad . \right.
\end{equation}
The solutions (\ref{rqs1}) may be rewritten as
\begin{equation}
\label{rqs2}
<a>_\eta = \left\{\sqrt{A_1\cos2(\eta - \eta_{01}) + B_1}
\quad \hfill\hbox{k = +1} \quad ,
\atop
\sqrt{A_2\cosh2(\eta - \eta_{02}) + B_2} \quad \hbox{k = -1}\quad .\right.
\end{equation}
The constants are given by
\begin{eqnarray}
(A_1)^2 = \biggr(18 - \frac{\sigma^2}{2} - \frac{p^2}{2}\biggl) + 36p^2 \quad &,& 
\quad (A_2)^2 = \biggr(18 + \frac{\sigma^2}{2} + \frac{p^2}{2}\biggl) - 36p^2 \quad ,\\
\tan2\eta_{01} = \frac{6p}{\frac{\sigma^2}{2} + \frac{p^2}{2} - 18} \quad &,&
\quad \tanh2\eta_{02} = - \frac{6p}{\frac{\sigma^2}{2} + \frac{p^2}{2} + 18} \quad ,\\
B_1 = \frac{\sigma^2}{2} + \frac{p^2}{2} + 18 \quad &,&
\quad B_2 = - \frac{\sigma^2}{2} - \frac{p^2}{2} + 18 \quad .
\end{eqnarray}
\par
Now, we will determine the classical analogue to this model.
After the introduction of the stiff repulsive fluid, the equations of motion read, in the conformal time gauge,
\begin{equation}
\biggr(\frac{a'}{a}\biggl)^2 = \frac{C_1}{a^2} - \frac{C_2}{a^4} -
k \quad ,
\end{equation}
which can be easily solved:
\begin{equation}
\label{rqs3}
a = \left\{\sqrt{A'_1\cos2(\eta - \eta_{01}) + \frac{C_1}{2}}
\quad \hfill\hbox{k = + 1} \quad ,
\atop
\sqrt{A'_2\cosh2(\eta - \eta_{02}) - \frac{C_1}{2}} \quad \hbox{k = - 1}\quad ,\right.
\end{equation}
where
\begin{equation}
A'_1 = \sqrt{\frac{{C_1}^2}{4} - C_2} \quad , \quad
A'_2 = \sqrt{\frac{{C_1}^2}{4} + C_2} \quad .
\end{equation}
The solutions (\ref{rqs3}) represent non-singular Universe and have the
same form as those obtained from the expectation value of the scale factor
in the quantum model.
Notice that the paremeter $p$, which leads to oscillations in the gaussian function,
is directly connected with the time phase $\eta_0$. The classical
analogue permit to give sense to some quantum parameters.
\par
In spite of the fact that the Wheeler-DeWitt equation for the spatially curved
radiative case is the same as the Schr\"odinger equation for the harmonic oscillator,
the final solutions exhibited here are different from the equivalent problem in
ordinary quantum mechanics. The reason for that relies again
on the fact that the dynamical variable here, the scale factor, must be positive definite,
a restriction which does not occur in the ordinary harmonic oscillator problem.
\par
The universality of repulsive classical fluid needed to reproduce the
quantum behaviour is more intringuing when we remark that it covers the
radiative curved case. Hence, thinking in terms of ordinary quantum mechanics, both
the free particle and the harmonic oscillator need the same repulsive
classical term in order to reproduce the quantum behaviour. Perhaps, other
curved cases (which are not reduced to a "free" particle or a harmonic oscillator problem)
may be treated in the same lines. But, the lack of simple closed expressions for
curved cases with $\alpha \neq 1/3$ in the quantum model makes this generalization
a hard task.

\section{Conclusions}

Quantum perfect fluid models in minisuperspace exhibit a dynamical variable,
connected with the matter degrees of freedom, which plays the
role of time. Hence, the Wheeler-DeWitt equation can
be reduced to a Schr\"odinger equation. For the flat case, solutions
are easily obtained for any ordinary fluid with a barotropic equation
of state $p = \alpha\rho$. However, wave packets can be constructed only
when $\alpha < 1$. For this case, the expectation value of the scale factor
reveals a singularity-free Universe which exhibits a bounce. The spreading of
the wave packet indicates a behaviour very similar to the free particle of
ordinary quantum mechanics.
\par
The main point of the present work is that these quantum perfect fluid models
admit a universal classical analogue such that the quantum
effects are reproduced by a repulsive fluid with an equation of state $p = \rho$.
This occurs also for the non-flat case with a radiative fluid. The existence
of this universal classical analogue, in the sense that the nature of the
repulsive fluid required to mimitize the quantum effects does not depend on
$\alpha$ and, perhaps, on $k$ indicates that the true nature of the quantum perfect fluid model
needs a deeper investigation.
\par
Indeed, the flat quantum cosmological perfect fluid model can be reduced, through
a canonical transformation, to the problem of a free particle in ordinary
quantum mechanics. The variable $x$ describing the free particle
is connected to the scale factor by the relation $x \propto a^{3(1 - \alpha)}$.
For $\alpha = 1/3$ we face a genuine free particle problem. But, for $\alpha \neq 1/3$,
the flat case has some similarity with that of a free particle, as the canonical transformation
employed in section 3 shows; there are also striking differences with
respect to the free particle problem and the similarity may not hide these differences.
The classical analogue is valid also for the
curved radiative case, which is equivalent to a harmonic oscillator problem. We may guess
that this analogue is valid to any value not only of $\alpha$ but also of $k$.
\par
Hence, results for quite different quantum cosmological models can be reproduced
classically using the same repulsive fluid in addition to the normal one which has
been employed in the quantization. This situation does not seem to occur in ordinary
quantum mechanics. In particular, it does not occur for the free particle and harmonic
oscillator problem. Here, this situation was assured due to the restriction of the
dynamical variable, the scale factor, to positive values, a restriction imposed by the
fact that we are treating a gravitational system. Hence, this analogous model seems to be
particular to the quantum cosmological model. 
\par
It can be argued that the question of equivalence between the quantum model
and the classical analogue has an obvious negative answer. Quantum mechanics
is a completly different framework compared with classical physics. The expectation
value of an observable in quantum mechanics may, in principle, be reproduced by a suitable classical model,
and in general this fact by itself has no deeper meaning. But here the situation is
more involved since it seems that quantum cosmology needs the notion of trajectory and, moreover, the modification introduced in the classical model in order to reproduce
the quantum one is universal, covering a large range of quite different models.
In our point of view, the universality of the term added to the classical model
in order to reproduce the results of the quantum model and the fact that this possibility
is typical of the gravity system represent surprising features of
the problem. For this reason, we think that the question of the formal
equivalence between the quantum cosmological perfect fluid model and the classical model
with a repulsive stiff matter fluid may hide a more profound meaning.
\newline
\vspace{0.5cm}
\newline
{\bf Acknowledgements:} We thank Gerard Cl\'ement for many enlightfull discussions and CNPq (Brazil) for partial financial support.

\end{document}